\documentclass[two column, amssymb, amsmath,nobibnotes,nofootinbib, aps, showpacs, pr]{revtex4}
\usepackage[dvips]{graphicx}
\begin{document}
\title{Investigation of weak itinerant ferromagnetism and critical behavior of  Y$_2$Ni$_7$}

\author{A.Bhattacharyya$^1$}
\author{Deepti Jain$^2$} 
\author{V. Ganesan$^2$} 
\author{S. Giri$^1$} 
\author{S. Majumdar$^1$} 
\email{sspsm2@iacs.res.in} 
\affiliation{$^1$Department of Solid State Physics, Indian Association for the
Cultivation of Science, 2A \& B Raja S. C. Mullick Road, Kolkata 700 032, INDIA} 
\affiliation{$^2$UGC-DAE Consortium for Scientific Research,
University Campus, Khandwa Road, Indore 452 017, INDIA}

\begin{abstract}

The weak itinerant ferromagnetic behavior of Y$_2$Ni$_7$ is investigated through magnetic, transport and calorimetric measurements. The low value of saturation moment, large Rhodes-Wohlfarth ratio, large value of the linear term in heat capacity and Fermi liquid like resistivity behavior with enhanced electron-electron scattering contribution establish firmly the weak itinerant ferromagetic nature of the sample. The critical exponents associated with the paramagnetic to ferromagnetic transition are also investigated from magnetization isotherms using  modified Arrott plot, Kouvel-Fisher plot and critical isotherm technique. The more accurate Kouvel-Fisher plot provides the critical exponents to be  $\beta$ = 0.306 , $\gamma$ = 1.401 and $\delta$ = 5.578. These values are markedly different from the mean field values and correspond closely to the two dimensional Ising spin system with long range spin spin interaction.  
  
\end{abstract}
\pacs{71.20.Be, 75.10.Lp, 75.40.Cx}
\maketitle

\section{Introduction}
Weak itinerant ferromagnets (WIFMs) are metallic magnetic systems characterized by very low saturation moment and low Curie temperature ($T_C$).~\cite{shimizu, kaul}  WIFMs attracted considerable attention due to their exotic ground state properties such as unusual magnetic excitation, superconductivity, quantum critical behavior, non-Fermi liquid (NFL) state etc.~\cite{saxena, cp, uhlarz}  The itinerant magnetism is based on the band theory of electrons and the magnetic moment arises from the exchange splitting of the band.~\cite{blanco} In case of WIFMs, the band splitting is extremely small, and as a result the high field moment per magnetic element  is only a fraction of a Bohr magneton. WIFMs lie very close to the magnetic non-magnetic phase boundary and often a small perturbation can give rise to a large change in the electronic and magnetic properties.

\par
 The conventional Stoner-Wohlfarth theory of band ferromagnetism~\cite{wohlf} based on Hartree-Fock mean field theory  is found to be inadequate to explain the various thermodynamical properties of WIFMs.   The theoretical model based on self consistent renormalization (SCR) approach  in presence of  spin fluctuations~\cite{moriya} is found to be more appropriate  describing the electronic and magnetic properties of such materials.  Among others, several transition metal based intermetallic compounds, such as ZrZn$_2$~\cite{zrzn2a, zrzn2b}, Sc$_3$In~\cite{sc3ina, sc3inb}, Ni$_3$Al~\cite{ni3ala, ni3alb},  MnSi~\cite{cp}, NbFe$_2$~\cite{nbfe2} etc. show weak itinerant ferromagnetism, which is  primarily connected to the delocalized nature of the $d$ electrons.  
 
 \par  
The title compound of the present work, Y$_2$Ni$_7$  is also a transition metal based WIFM.  Y-Ni solid solutions show interesting magnetic properties depending upon the ratio of Y and Ni.\cite{dg1, dg2} The stoichiometric compounds YNi$_{17}$ and YNi$_{15}$ show ferromagnetic ground state. However, saturation moment and $T_C$ decrease  as the Ni concentration is further lowered and eventually YNi$_5$ becomes a  paramagnetic material. With further decrease  in Ni concentration, ferromagnetism reappears and the compounds such as Y$_2$Ni$_7$ and YNi$_3$ show weak itinerant ferromagnetism.~\cite{nishi, naka, tazuke} Finally,  ferromagnetic state  disappears in case of  YNi$_2$.  Evidently, Y$_2$Ni$_7$ remains close to the magnetic  and non-magnetic phase boundary. 
 
 \par
 The weak itinerant character of ferromagnetism in Y$_2$Ni$_7$ is apparent  from the low saturation moment and low $T_C$ ($\sim$ 50 K) as reported in previous studies.~\cite{nishi,rb}  However, a comprehensive investigation of the ground state magnetic and electronic properties of the material is lacking. Such investigation is necessary  to understand the low lying excitation in the system, where magnetic fluctuations play a very crucial role in determining the magnetic state. Another important aspect associated with the low moment ferromagnet like Y$_2$Ni$_7$ is the nature of the paramagnetic (PM) to ferromagnetic (FM) phase transition. For that, one needs to investigate the critical region and the corresponding critical exponents. This can provide us the order, the universality class and the effective dimensionality  of the phase transition. 
 
 \par
The present paper is organized in the following manner. Firstly, a thorough investigations have been presented based on the magnetic, transport and heat capacity measurements.  Later we have provided the criticality and scaling behavior of Y$_2$Ni$_7$ around $T_C$ using modified Arrott plot, Kouvel-Fisher plot and critical isotherm method.

\begin{figure}[t]
\vskip 0.4 cm
\centering
\includegraphics[width = 8.5 cm]{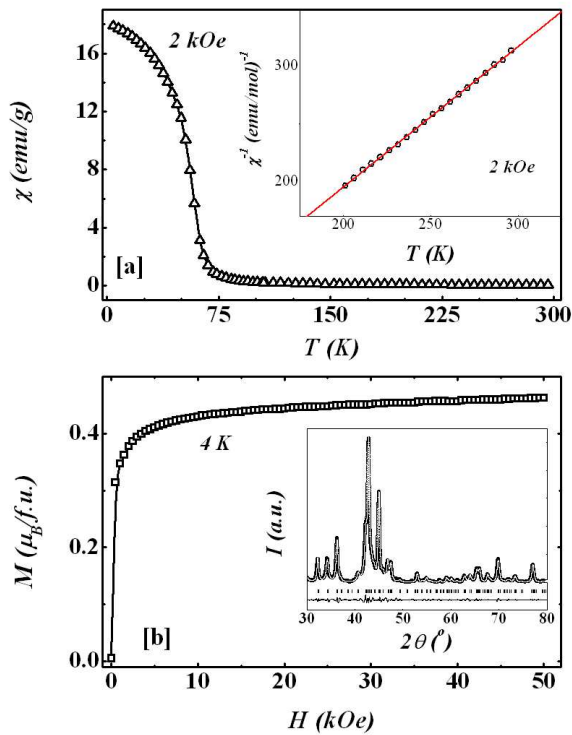}
\caption {(Color online)(a) Temperature dependence of dc magnetic susceptibility ($\chi = M/H$) measured in zero-field-cooled condition in  presence of an applied magnetic field of 2 kOe. Inset shows the inverse dc magnetic susceptibility measured  above 200 K. (b) shows the isothermal field dependence of magnetization at 4 K. The inset in (b) represents the XRD pattern of the sample along with the Rietveld refinement data. }
\end{figure}

\begin{figure}[t]
\vskip 0.4 cm
\centering
\includegraphics[width = 9 cm]{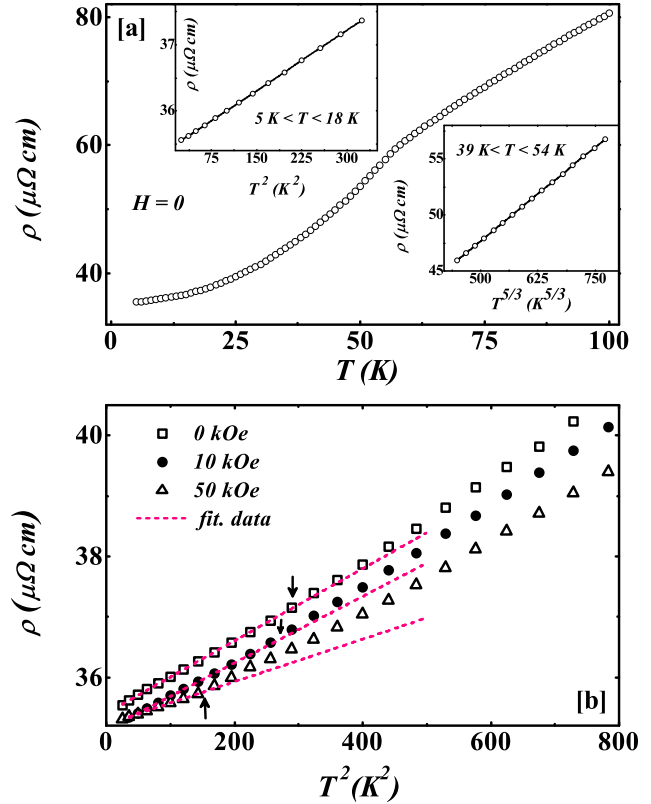}
\caption {(Color online)(a) Resistivity ($\rho$) as a function of temperature for Y$_2$Ni$_7$.  The insets show the $T^2$ and $T^{5/3}$ variation of $\rho$ at low temperature (5-18 K) and just below (39-54 K) the magnetic transition ($T_C$) respectively. (b) shows the low temperature resistivity as a function of square of the temperature for different applied fields. The dashed  lines are the  linear fit to the data.   }
\end{figure}

\begin{figure}[t]
\vskip 0.4 cm
\centering
\includegraphics[width = 8.5 cm]{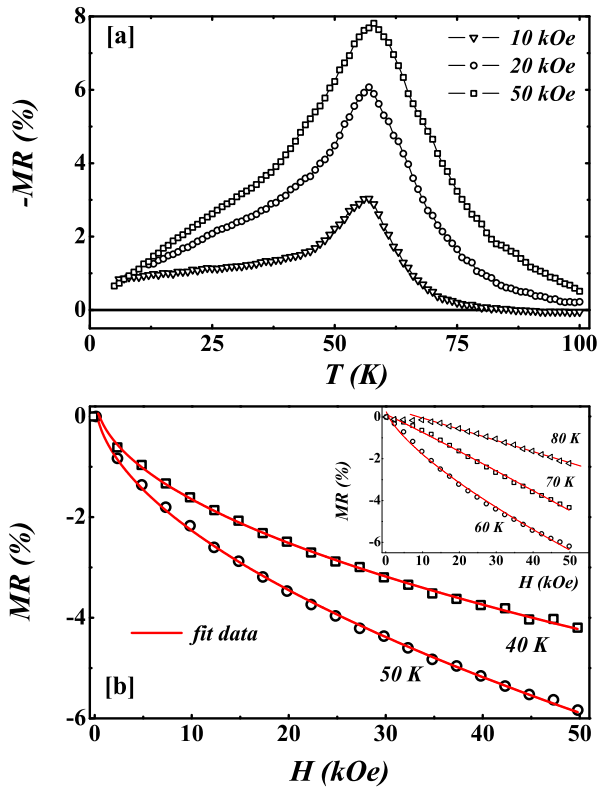}
\caption {(Color online)(a)  shows the temperature dependence of magnetoresistance (MR = [$\rho(H)-\rho(0)$]/$\rho(0)$) for different applied fields.  (b)  The main panel shows the variation of MR with field at two selected temperatures below $T_C$. The solid line shows the fit to the data with equation MR = $aH^{0.5}$. The inset shows the MR as a function of field  at few temperatures above $T_C$ along with the fitted curves (solid lines).}
\end{figure}

\begin{figure}[t]
\vskip 0.4 cm
\centering
\includegraphics[width = 8.5 cm]{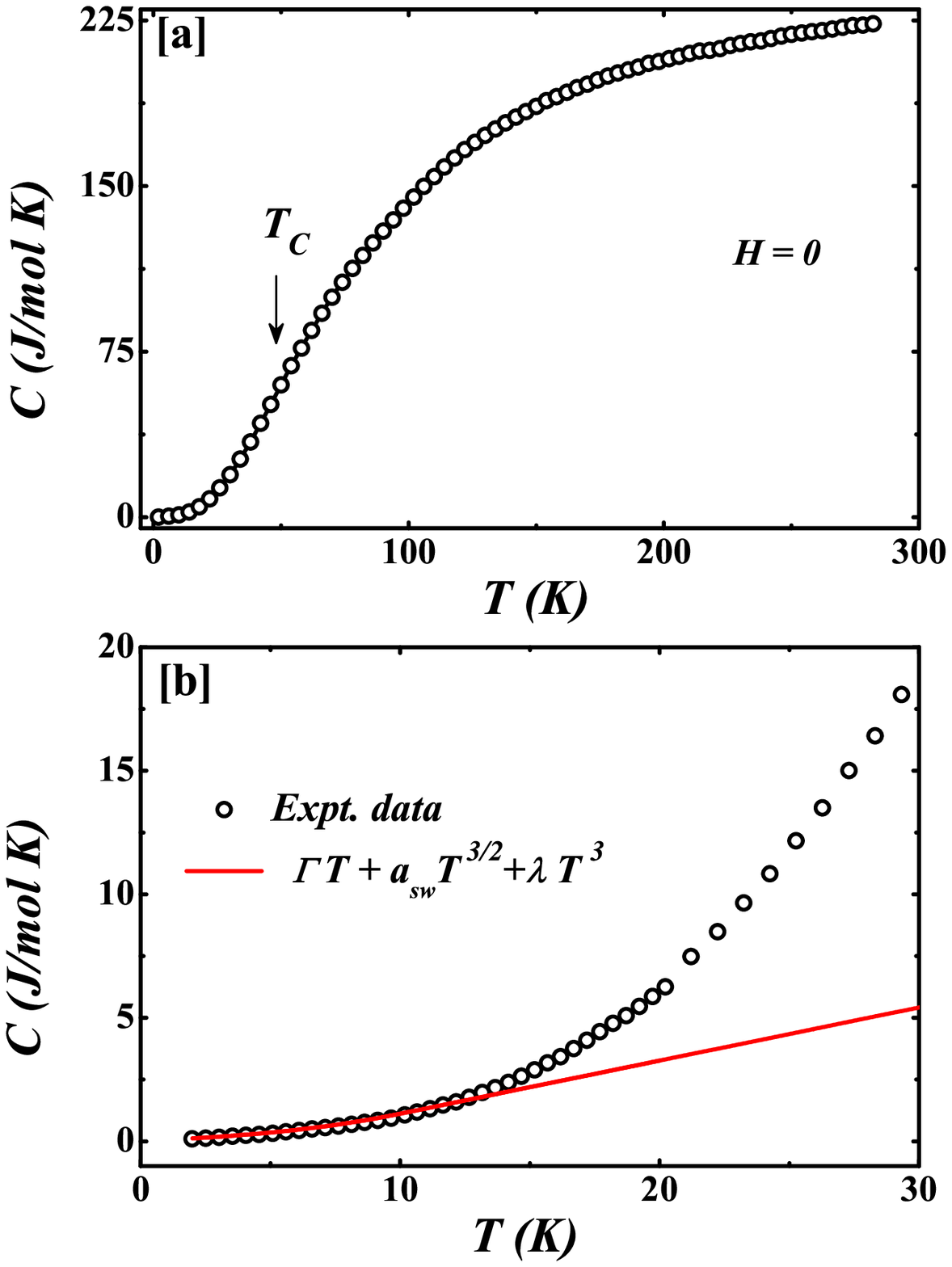}
\caption {(Color online)(a) shows the heat capacity as a function of temperature .  (b) shows the  low temperature heat capacity with solid line being fit to the data  with a formula comprising electronic, lattice and spin wave contributions of heat capacity.}
\end{figure}

\section{Experimental Details}
The bulk polycrystalline sample of  Y$_2$Ni$_7$  for the present investigation was prepared by argon arc melting followed by annealing at 1000$^{\circ}$C for 500 h. The x-ray powder diffraction pattern (using Cu K$_{\alpha}$) were collected  using a SEIFERT XRD3000P diffractometer (Cu K$_{\alpha}$ radiation, 2$\theta$ range from 30$^{\circ}$ to 80$^{\circ}$ with step size 0.02$^{\circ}$).  The collected powder patterns have been used for Rietveld refinement (see inset of fig. 1 (b)) using the GSAS software package.~\cite{gsas} The analysis shows that  the sample was formed in single phase with rhombohedral Gd$_{2}$Co$_7$ type crystal structure (space group: $R\overline{3}m$) with lattice parameters $a$ = 4.952 \AA, $c$ = 36.230 \AA,  and $V$ = 888.44 \AA$^3$, which  match well with previous result.~\cite{bu} 

\par
Magnetization ($M$) measurements on Y$_2$Ni$_7$ sample was carried out on a Quantum Design SQUID magnetometer (MPMS 6, Evercool model). The resistivity ($\rho$) and magnetoresistance (MR) were measured on a cryogen-free high magnetic field system from Cryogenic Ltd., UK. The heat capacity ($C$) was recorded on a Qantum Design Physical Property Measurement System down to 2 K.

\section{Magnetization, transport and thermodynamical studies}
\subsection{Magnetization}
The temperature ($T$) variation of the dc magnetic susceptibility ($\chi = M/H$, where $H$ is the applied magnetic field) measured in zero field cooled  condition in presence of $H$ = 2 kOe is shown in fig. 1 (a).  $\chi$ shows a sharp rise  below 70 K with decreasing $T$. This corresponds to the PM to FM transition in the sample and the associated $T_C$ is found to be around 53 K. This value of $T_C$ has been calculated from the first $T$ derivative of $\chi$ (not shown here). Above 200 K,  $\chi^{-1}(T)$ varies linearly with $T$ (see inset of fig. 1 (a)), which signifies the validity of Curie-Weiss law at high-$T$.  Linear fit to the high-$T$ part of $\chi^{-1}(T)$ {\it vs.} $T$ data gives the paramagnetic effective moment $p_{eff}$= 0.93 $\mu_B$/Ni and paramagnetic Curie temperature $\theta_p$ = 40 K. These values are close to the previously reported results.~\cite{bu1}

\par
Fig. 1(b) shows the $M$ versus $H$ isotherm recorded at 4 K.  We observe FM like behaviour with steep rise at low $H$ and a sluggish increase at higher fields. Till $H$ = 50 kOe, $M$ does not show complete saturation. This lack of saturation is presumably connected to the itinerant character of the ferromagnetism.~\cite{kaul}  The  magnetic moment for $H$ = 50 kOe  is found to be 0.06 $\mu_B$/Ni, which is  one order of magnitude smaller than the {\it per atom} moment in metallic Ni  (0.64 $\mu_B$/Ni).~\cite{ni} Such small value of  moment in Y$_2$Ni$_7$ is an indication of the very weak FM nature of the sample. The observed moment is  comparable with the results reported  in previous studies on Y$_2$Ni$_7$.~\cite{bu1, nishi}   It is to be noted that there is  no hysteresis in the field increasing and decreasing legs of the  $M-H$ isotherm, indicating a very soft FM character of the sample.

\par
The itinerant ferromagnetism is often  characterized by the Rhodes-Wolhfarth ratio, which is defined as RWR = $p_c/p_{sat}$. Here  $p_c$ is related to $p_{eff}$  by the relation  $p_{eff}$ = $p_c$($p_c$ +2), while $p_{sat}$ is the saturation moment per magnetic atom  at the temperature of interest.~\cite{wol} For a localized system,  the value of RWR should be close to 1, while it diverges for itinerant ferromagnets.  For Y$_2$Ni$_7$, $p_{sat}$ has been calculated from the $M-H$ isotherm at 4 K and the resulting RWR is found to be 6.17. This value is much larger than unity and provides support for the itinerant character of ferromagnetism in Y$_2$Ni$_7$. It is to be noted that  for a  typical WIFM such as ZrZn$_2$, the RWR is close to 5.4.~\cite{ak}

\begin{figure}[t]
\vskip 0.4 cm
\centering
\includegraphics[width = 8.5 cm]{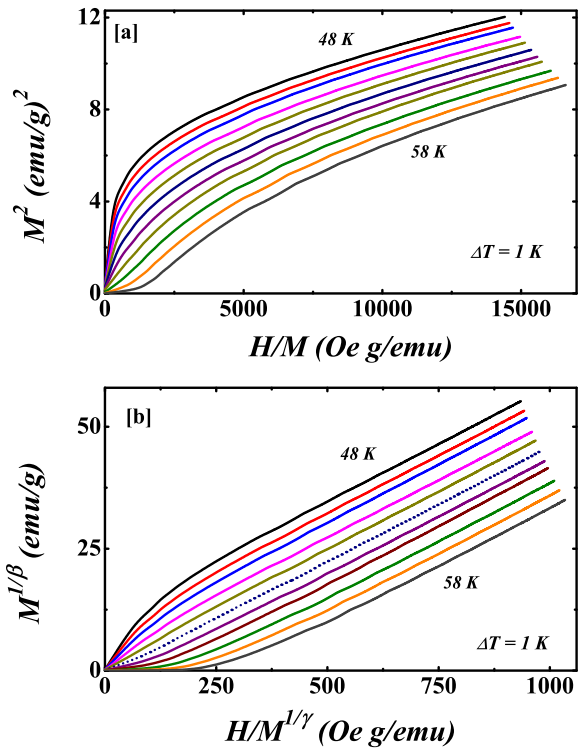}
\caption {(Color online) (a) shows several $M^2$ vs $H/M$ isotherms (Arrott plot)  of Y$_2$Ni$_7$  around $T_C$ ($\approx$ 53 K) with temperature interval $\Delta T$ = 1 K . (b) represents modified Arrott plot ($M^{1/ \beta}$ vs $(H/M)^{1/\gamma}$) with $\beta$ = 0.31 and $\gamma$ = 1.40 . At 53 K, the modified Arrott plot  pass through the origin indicating the proximity of  Curie temperature.}
\end{figure}

\subsection{Resistivity and magnetoresistance}

$T$ dependence of $\rho$ for Y$_2$Ni$_7$ is shown in fig. 2 (a). Clear signature of change in slope is observable around $T_C \approx$ 53 K.  At low-$T$ (below $\sim$ 18 K, in the FM state), $\rho(T)$ shows a well defined $T^2$ dependence ($\rho(T) = \rho_0 + BT^2$) as evident from the upper inset of fig. 2(a). Such  $T^2$ dependence is the typical Fermi liquid behavior arising from the electron-electron scattering which generally does not involve spin-flip process.  Such Fermi liquid type behavior is observed in case of many simple metals at low-$T$.~\cite{fli1,nfl} The SCR theory which incorporates spin fluctuations in WIFM also predicts a $T^2$ dependence of $\rho$ well below $T_C$.  However, this $T^2$ dependence has different origin than the simple electron-electron scattering  and it is connected to the  {\it spin-flip} scattering of $s$ electrons by the spin density fluctuations of the $d$ electrons via $s-d$ exchange interaction.~\cite{ueda1, hertel} Notably , the coefficient $B$ of the $T^2$ term is found to be rather large  in case of Y$_2$Ni$_7$ with  its value being  5.04 $\times$10$^{-9} \Omega$ cm$^{-1}$K$^{-2}$. This is about two order of magnitude higher than the typical FM metal such as Ni and Fe ($\sim$ 10$^{-11} \Omega$cm$^{-1}$ K$^{-2}$). Similar enhanced value of $B$ was observed in prototypical WIFM such as ZrZn$_2$  or Ni$_3$Al.~\cite{ogawa1, ogawa2} 

\par
 The spin fluctuation theory  predicts  $T^{5/3}$ dependence of $\rho$ for WIFM just below $T_C$.~\cite{ueda1} The lower inset in fig. 2 (a) shows $\rho$ as a function of $T^{5/3}$ just below  $T_C$  and  the  linear nature of the  curve  indicates that $\rho$ varies as $T^{5/3}$, which is consistent with the prediction of SCR model.
\par
We also  investigated the $\rho (T)$ behavior of the sample under  $H$. The $\rho$  {\it vs.} $T^2$ data at low temperature are shown here for different values of  $H$ (fig. 2(b)). It is  evident that the  temperature range where we can observe  $T^2$ dependence of $\rho$  becomes narrower with increasing $H$. The Fermi liquid type $T^2$ dependence is observed below 18 K in zero field, however in 10 kOe and 50 kOe of fields,  it is only visible below  15 K and 12 K respectively . Such suppression of Fermi liquid like state with $H$  indicates  some field-induced  change in the $s-d$ scattering process.~\cite{fli1,fli2}  We have fitted the $T^2$ dependent part of $\rho(T)$ to the formula $\rho(T) = \rho_0 + BT^2$ and it has been observed that $B$ decreases with increasing $H$. This is due to the fact that the applied field tends to quench the spin fluctuations.~\cite{ueda2}  

\par

We have calculated the  magnetoresistance (MR = [$\rho(H)-\rho(0)]/\rho(0)$) of the sample from the $\rho$ versus $T$ data recorded under different values of $H$. The $T$ variation of -MR is plotted in fig. 3(a) between 5 and  100 K.  MR is found to be negative over a  wide temperature region both above and below $T_C$ with  -MR versus $T$ data showing a peak close to $T_C$. At 50 kOe , the maximum value of MR  is found to be about -8\% around 58 K.  The  observed negative MR is  likely to be  associated with the suppression of spin fluctuations under $H$.~\cite{ikeda} If the Zeeman splitting energy corresponding  to the applied magnetic field between spin up and spin down states is comparable or larger than the spin fluctuation energy, the inelastic spin flip scattering probability decreases leading to the decrease in $\rho$.  Observation of negative MR well over $T_C$ indicates that the spin fluctuations exist even at higher temperatures for Y$_2$Ni$_7$.  Ueda~\cite{ueda2}  studied the effect of magnetic field on the spin fluctuations in WIFM based on the SCR theory. The results indicate the  spin fluctuations related negative MR to be present  both above and below $T_C$. The MR calculated from SCR theory is found to be negative with its maximum value at $T_C$. It also predicts that  the range above $T_C$, over which negative MR is observed, increases with increasing $H$. Very similar effect is observed in case of Y$_2$Ni$_7$ with the region of negative MR above $T_C$ being widened with increasing $H$.

\par
Fig. 3(b) shows the isothermal MR versus $H$ data above (inset)  and below (main panel) $T_C$. In both the regions,  MR is found to follow power law of the type MR $\sim H^m$.  The exponent $m$ is found to be 0.5 below $T_C$ as evident from the 40 K and 50 K data. On the other hand just above $T_C$ (60 K), $m$ shows slightly enhanced value of 0.7.  On further increase of $T$, MR varies almost linearly with $H$ (except for the low field part)  as apparent from the 70 K and 80 K isotherms shown in the inset.   
\begin{table}
\begin{center}
\caption{Basic characteristic of Y$_2$Ni$_7$ together with some itinerant weak ferromagnets.~\cite{ak}}
\begin{tabular}{lccccccccccccccc}
\hline
   && Y$_2$Ni$_7$  && ZrZn$_2$  && InSc$_3$ && Y$_4$Co$_3$   \\   
\hline
\hline
$\theta_p$(K)  && 40   && 33 && 8 && 14    \\
$T_C$(K)  && 53   && 21 && 6 && 5    \\ 
$p_c$($\mu_B$/at.)  && 0.37   && 0.65 && 0.26 && 0.14        \\    
$p_{sat}$($\mu_B$/at.)  && 0.06   && 0.12 && 0.045 && 0.012     \\ 
RWR = $p_c$/$p_{sat}$  && 6.17   && 5.4 && 5.78 && 11.5       \\ 
$\Gamma$ (mJ/mol K$^2$) && 52.3 &&45&&12 && 3.45 \\
\hline 
\hline
\end{tabular}
\end{center}
\end{table}

\subsection{Heat Capacity}
The $C$ versus $T$ data  of Y$_2$Ni$_7$ from 2 K to room temperature is shown in fig. 4 (a). Apparently no anomaly is observed near $T_C$ in the $C(T)$ plot.  This is due to the fact that the low moment ordering associated with the WIFM has a small contribution as compared to the lattice  and electronic components of $C$. We have  carefully looked the low $T$ behavior of $C$ as shown in fig. 4 (b).  At $T \ll \Theta$ ($\Theta$ = Debye temperature), the lattice part of the heat capacity $C_{debye}$  have a $T^3$ dependence. The other contributions for $C$ will be the spin wave term of the ordered FM state and a linear term corresponding to the electronic heat capacity. The solid line in fig. 4(b)  represents a fit to the data with the contributions $C = \Gamma T + a_{sw}T^{3/2} + \lambda T^3$, where the terms respectively denote the electronic heat capacity, spin wave contribution and the low-$T$  lattice contribution. The best fit is obtained for $\Gamma$ = 52.3 mJ mol$^{-1}$ K$^{-2}$, $a_{sw}$ = 2.61 mJ mol$^{-1}$ K$^{-5/2}$ and $\lambda$ =0.506 mJ mol$^{-1}$ K$^{-4}$. $\lambda$ is related to the Debye temperature $\Theta$ as $\lambda = \frac{12}{5}\pi^4\frac{pR}{\Theta^3}$, where $p$ is the number of atoms per formula,  and $R$ is the universal gas constant. Using this relation, we get the  value of $\Theta$ to be 306 K. The interesting point to be noted here is the enhanced value of $\Gamma$, which is  otherwise  close to   1 mJ mol$^{-1}$ K$^{-2}$ for simple metal like Cu. Such enhancement  is clearly related to the strong spin fluctuation in WIFM as described by Moriya.~\cite{moriya} Enhanced $\Gamma$ has also been observed in case of other WIFM, for example, it is close to 45  mJ mol$^{-1}$ K$^{-2}$ in ZrZn$_2$.~\cite{zrzn2b} The spin wave contribution for Y$_2$Ni$_7$ is also rather weak, which is also a characteristic feature of a WIFM.~\cite{heusler}

\par
A comparison of the electronic and magnetic properties of Y$_2$Ni$_7$ with few other well known WIFM is depicted in table I. The observed parameters characterizing the WIFM behavior in Y$_2$Ni$_7$ are close to the values observed in other materials.

\begin{figure}[t]
\vskip 0.4 cm
\centering
\includegraphics[width = 8 cm]{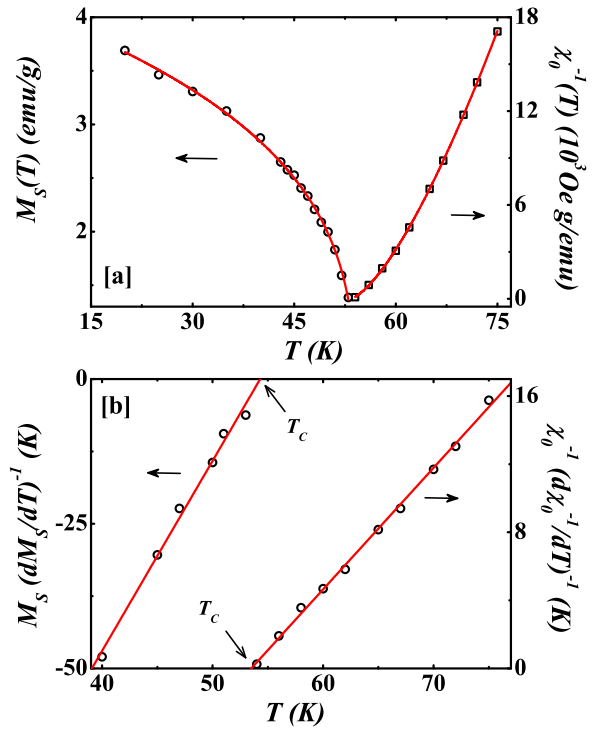}
\caption {(Color online) (a) Temperature dependence of spontaneous magnetization $M_S $ (left axis) and inverse initial susceptibility $\chi^{-1}_{0}$ (right axis) which are obtained from the high-field extrapolation of modified Arrott plot (fig. 5 (b)). The solid lines represent the fit to the data by eqns. 1 and 2. (b)  Kouvel-Fisher plot of spontaneous magnetization $M_S$ (left axis) and inverse initial susceptibility $\chi^{-1}_{0}$ (right axis) for Y$_2$Ni$_7$. Straight lines are the  linear fit to the data.}
\end{figure}

\section{Scaling and Critical exponents}

\subsection{Modified Arrott plot}
For a second order phase transition from a PM  to FM phase, the critical behavior near $T_C$ is characterized by a set of critical exponents, namely $\beta$, $\gamma$ and $\delta$, which are respectively associated with the spontaneous magnetization ($M_S$), initial susceptibility ($\chi_0 = \lim_{H \to 0} M/H$) and magnetization isotherm ($M-H$). The scaling hypothesis suggests following power law relations near the critical region: ~\cite{stanley} 

\begin{eqnarray}
M_S(T)&=&M_0 {|\epsilon|}^{\beta},  \epsilon < 0 ,  T < T_C \\
\chi^{-1}_{0} (T)& =& G(\epsilon)^{\gamma} , \epsilon > 0 ,  T > T_C  \\ 
M &=& XH^{1/\delta} ,  \epsilon = 0 ,  T = T_C  \\
\nonumber 
\end{eqnarray}

Where $\epsilon$ = $(T-T_C)/T_C$ is the reduced temperature. $M_0$, $G$ and $X$ are the critical amplitudes. 

\par

In order to determine the  critical exponents of Y$_2$Ni$_7$, we  recorded several $M$ versus $H$ isotherms. In fig. 5(a)  we plot  $M^2$  as a function of  $H/M$ at different temperatures around $T_C$, which is commonly known as Arrott plot.~\cite{arrott} In the mean field theory, one would expect   $M^2$  {\it vs.}  $H/M$ isotherms to be series of parallel straight lines around  $T_C$ and the line would pass through the origin at $T = T_C$.  The main observations from the Arrott plot are: (i) The curves are non-linear even  at high field, which rules out the possibility of a mean field model for the phase transition at $T_C$; and (ii) the plots show downward curvature, {\it i.e., } the slope of the curves  is always positive.  According to the condition suggested by Banerjee,~\cite{bn} the positive slope indicates that the phase transition at $T_C$ is second order in nature.  The occurrence of second order phase transition allows us to investigate the critical behavior of  Y$_2$Ni$_7$  on the the basis of exponents as described in eqns. 1-3. 

\begin{figure}[t]
\vskip 0.4 cm
\centering
\includegraphics[width = 8.5 cm]{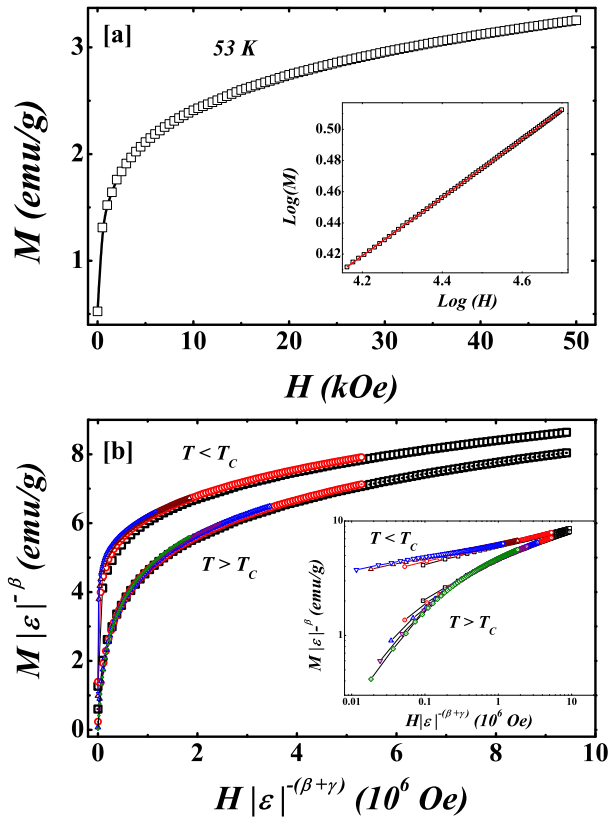}
\caption {(Color online) (a) Magnetization isotherm collected at  53 K for Y$_2$Ni$_7$. Inset shows the same plot in log-log scale and the straight line is the linear fit following eqn. 3.  (b) shows the reduced magnetization ($M/\epsilon^{\beta}$)   plotted against the reduced field ($H/\epsilon^{\beta + \gamma}$). The plot shows all the data collapse into two separate branches: one below $T_C$ and another above $T_C$. Inset shows the same plot in the log-log scale.}
\end{figure}
\par
Considering  non-mean field like behavior of the second order phase transition in Y$_2$Ni$_7$, we have  used more generalized modified Arrott plot techniques. This is based on the Arrott-Noakes equation of state in the critical region:~\cite{an}
 
\begin{eqnarray}
(H/M)^{1/\gamma} &=& a(T-T_C)/T + bM^{1/\beta}
\end{eqnarray}
where $a$ and $b$ are constants.

\par
In modified Arrott plot $M^{1/\beta}$ is plotted against  $(H/M)^{1/\gamma}$ for suitable choice of the exponents of $\beta$ and $\gamma$. The proper choice of $\beta$ and $\gamma$ will produce modified Arrott plot, where the curves are parallel to each other at least in the high field region. We checked this for $\beta$ and $\gamma$ values of 3D Heisenberg model, 3D Ising model etc., but none of them provide parallel straight lines in the modified Arrott plot.  Hence for the proper choice of the exponents, we varied $\beta$ and $\gamma$ over a wide range staring from the above models.  After comparing large number \{$\beta, \gamma$\}, it has been found the best parallel sets  of straight line for modified Arrott plot occurs for $\beta$ = 0.31 and $\gamma$ = 1.4. Fig. 5(b) shows such plot of Y$_2$Ni$_7$ at selected temperatures around $T_C$. We can also calculate the value of $\delta$ using Widom scaling relation, $\delta = 1+\gamma/\beta$ and the corresponding value of $\delta$ is found to be 5.578.

\par
Modified Arrott plot  provides a good opportunity to calculate the values of $M_S$ and $\chi_0$. Linear extrapolation of the high field straight line portion of the isotherm provides the value of ($M_S)^{1/\beta}$ and ($\chi^{-1}_{0})^{1/\gamma}$ as an intercept on the ($M)^{1/\beta}$ and ($H/M)^{1/\gamma}$ axes respectively. Now using the values of $\beta$ and $\gamma$ of the modifies Arrott plot, one can calculate $M_S$ and $\chi^{-1}_{0}$.   

The $T$ dependence of $M_S$ and $\chi^{-1}_0$, obtained from the intercepts of the modified Arrott plot, is  plotted in fig. 6 (a). The variation of these quantities in the critical region satisfies the behavior expected  from scaling laws (eqns. 1 and 2). We have fitted the $M_S(T)$ and $\chi^{-1}_0(T)$ curves with eqns. 1 and 2 respectively. This provide a new set of values of $\beta$ and $\gamma$. The value of $\beta$ obtained from $M_S(T)$ data is 0.292 ($\pm$ 0.006) while the $\gamma$ value from  $\chi^{-1}_0(T)$ data is 1.44 ($\pm$ 0.02). 

\subsection{Kouvel-Fisher method}
A more accurate method to determine the exponents $\beta$ and $\gamma$ is by the Kouvel-Fisher method.~\cite{ni}  It is based on the equations:
\begin{eqnarray}
\frac{M_S}{dM_S/dT} = \frac{T-T_C}{\beta} \\
\frac{\chi_0^{-1}}{d\chi_0^{-1}/dT} =\frac{T-T_C}{\gamma} 
\end{eqnarray}

Therefore, the $T$ variation  of  $M_S (dM_S/dT)^{-1}$ and $\chi^{-1}_{0}/(d\chi^{-1}_{0}/dT)$  yields respectively straight lines with slopes $1/\beta$ and $1/\gamma$  along with the  intercepts on the $T$ axis giving the values of $T_C$. The $T$ variation of such quantities is shown  in fig. 6 (b) and the resulting  linear curves confirm the applicability of the Kouvel-Fisher method for the present sample.  By linear fit of the left curve ($M_S (dM_S/dT)^{-1}$  versus $T$), we have obtained $\beta$ =0.306 ($\pm$ 0.002) and $T_C$ = 54.42 K. On the other hand linear fit of the right curve ($\chi^{-1}_{0}/(d\chi^{-1}_{0}/dT)$ versus $T$) provides $\gamma$ = 1.401($\pm$ 0.02) and $T_C$ = 53.58 K. These values are quite close to that obtained from modified Arrott plot method.

\par
The value of $\delta$ can also be independently calculated from $M-H$ isotherms by using eqn. 3. Fig. 7 (a) represent the $M-H$ isotherm at 53 K (the closest temperature to the $T_C$) along with $\log M$ versus $\log H$ plot in the inset.  According to eqn. 3, the $\log M$ versus $\log H$ plot should be straight line with slope to be 1/$\delta$ at $T_C$. We have calculated the value of $\delta$ from the log-log plot and it is found to be 5.35. This value is also quite close to that value obtained from modified Arrott plot and Widom scaling relation.

 \subsection{Scaling theory}
 The critical exponents obtained from different methods are found to be close to each other.  However, it is important to check the reliability of the values through the scaling hypothesis. According to the hypothesis~\cite{stanley, kaul2} $M(H, \epsilon)$ is a universal function of $T$ and $H$ and there exists a reduced equation of state of the form:
 
\begin{eqnarray}
M(H,\epsilon) &=& \epsilon^{\beta}f_{\pm} (H/\epsilon^{\beta+\gamma}) 
\end{eqnarray}

\par
where the functions  $f_{+}$ and $f_{-}$  are  for $T>T_C$ and  $T<T_C$ respectively. This equation implies that experimentally observed  $M/\epsilon^{\beta}$ versus $H/\epsilon^{\beta + \gamma}$ should collapse on two different curves, one for temperature below $T_C$ and the other for above $T_C$.  Such scaling would be realized if one chooses right values of the  $\beta$, $\gamma$ and  the $T_C$.

\par
In order to check this criterion for Y$_2$Ni$_7$, we have plotted  $M/\epsilon^{\beta}$  as a function of $H/\epsilon^{\beta + \gamma}$ (see fig. 7(b))  in the critical region with the values of $\beta$ and $\gamma$ obtained from Kouvel-Fisher method.  It is to be noted that all the curves  converges into two branches depending upon  $T > T_C$ or $T<T_C$.  The inset of fig. 7(b) shows the same plot in the log-log scale for better clarity. This  shows that  the scaling hypothesis  is obeyed over a wide range of $T$ and $H$, and therefore the calculated critical exponents are meaningful.

\begin{table}
\begin{center}
\caption{Critical exponents  $\beta$, $\gamma$ and $\delta$ obtained from modified Arrott plot (MAP), Kouvel-Fisher (KF) plot and critical isotherm (CI) for Y$_2$Ni$_7$ along with theoretical values for various model.~\cite{kaul2}}
\begin{tabular}{lcccccccccccc}
\hline
    && $\beta$  && $\gamma$ && $\delta$  \\   
\hline
\hline
Y$_2$Ni$_7$ (MAP)   && 0.31 && 1.40  && 5.52         \\ 

Y$_2$Ni$_7$ (KF)   && 0.306&& 1.401 && 5.578         \\

Y$_2$Ni$_7$ (CI)   && -&& -  && 5.35         \\

Mean Field Model     && 0.5 && 1.00 && 3      \\    

3D Heisenberg Model     && 0.365 && 1.386 && 4.8       \\

3D Ising Model     && 0.325  && 1.241  && 4.82      \\ 

\hline 
\hline
\end{tabular}
\end{center}
\end{table}

\section{Discussion} 
The present work aims to  study of the magnetic and electronic properties of Y$_2$Ni$_7$ based on the magnetization, transport and calorimetric measurements. The signature of PM to FM transition in Y$_2$Ni$_7$ is found to be relatively weak in our $M(T)$, $\rho(T)$ and $C(T)$ data indicating the weak ferromagnetism in the sample. This is  supported by the low value of the saturation moment as evident from the  $M(H)$ isotherms. The ground state physical properties  indicate the WIFM character of the sample, which is particularly evident from the  large value of the coefficient of $T^2$ term in $\rho$ and large coefficient of the linear term in $C$. Such enhancement is connected to the spin fluctuations present in  the itinerant ferromagnet.

\par
In Y$_2$Ni$_7$,  spin fluctuations exist  well above $T_C$. Notably, such behavior is supported by the theoretical calculation based on SCR model.~\cite{ueda2} In our experimental data, we observe negative MR  
at temperature as high as 100 K, which is about twice the $T_C$ of the sample.   It was shown that the MR arising from spin fluctuations in a weakly FM material  should follow a linear field dependence {\it i.e.,} (MR)$_{SF} \sim -H$. Below $T_C$ we observe MR to vary as $-H^{0.5}$ rather than a linear field dependence. Although MR does not  show linear variation with $H$  below $T_C$,  we observe linear $H$ dependence of MR at high field for $T >$ 60 K.
  
 \par 
The $T$ dependence of $\rho$ and $C$ shows typical Fermi liquid like behavior with enhanced value of the coefficients of $T^2$ and $T$ terms respectively.  It is to be noted that their ratio $B/\Gamma^2$ is an important parameter to  ascertain the Fermi liquid state in a metal.~\cite{kadowaki, rice} The coefficient $B$ in $\rho$ arises from  the electron-electron scattering and it is found to be proportional to the square of the effective mass of  the conduction electrons. On the other hand the electronic specific heat coefficient is linearly proportional to the effective mass. Therefore, within a class of materials obeying renormalized band picture, the ratio should  have universal value. For heavy fermion metals, the ratio is found to be close to 1.0$\times$10$^{-5} \Omega$ cm mole$^2$ K$^2$ J$^{-2}$.~\cite{kadowaki}  However, for transition metals, the ratio has an average value of 0.9 $ \times$10$^{-6}\Omega$ cm mole$^2$ K$^2$ J$^{-2}$, which is one order of magnitude lesser than that of the heavy fermions. We have calculated  the ratio for Y$_2$Ni$_7$ and it turns out to be 1.8 $ \times$10$^{-6} \Omega$ cm mole$^2$ K$^2$ J$^{-2}$, which is pretty close to the value found in case of transition metals.  Although the coefficients $B$ and $\Gamma$ are much enhanced in case of Y$_2$Ni$_7$ due to spin fluctuations, the ratio remains the same. This indicates that the spin fluctuations in Y$_2$Ni$_7$ can be well accounted by the {\it renormalized electronic band parameters}.

\par
One interesting observation is the effect of $H$ on the $T^2$ dependent part of the $\rho(T)$ data. The temperature window over which $\rho \sim T^2$, diminishes with increasing $H$. The SCR theory, contrary to our experimental result, actually predicts the broadening of the $T^2$ dependent region with $H$,~\cite{ueda2} and similar behavior was observed experimentally.~\cite{heusler} This contradictory result in Y$_2$Ni$_7$ might be an  indication of field induced  instability in the Fermi liquid ground state of Y$_2$Ni$_7$.    It is to be noted that the sample remains FM in high fields and $T_C$ does not decrease with increasing $H$. Therefore, the emergence of NFL like state with $H$ can not be attributed to the  fact that the system is nearing to a  quantum critical point.~\cite{nfl} It is to be noted that   ZrZn$_2$ shows NFL like resistivity behavior at the ambient condition in the FM state.~\cite{zrzn2b} Coexistence of FM and NFL states has  been observed in the alloys URu$_{2-x}$Re$_x$Si$_2$ and it is found to be feasible if we consider  the material to be disordered and anisotropic leading to the formation of FM clusters.~\cite{bauer} Such scenario can not be ruled out in case of Y$_2$Ni$_7$. On the other hand, electronic structure calculations indicates  a weak peak in the density of states (DOS) near the Fermi level of Y$_2$Ni$_7$, which is found to be responsible for the WIFM character.~\cite{band} Any subtle change in the DOS with $H$ may also result in reducing the $T^2$ dependent range of $\rho$.

\par
The critical exponent calculated from the modified Arrott plot, Kouvel-Fisher plot and the Critical isotherm method has been depicted in table II. It is evident that the values calculated from different techniques are quite close. Among other methods, the Kouvel-Fisher plot can provide the most accurate values of the critical exponents. The exponents from Kouvel-Fisher method show scaling behavior, where the scaled equation of state (eqn. 7) produces two curves for the state below and above $T_C$ (see fig. 7 (b)). This proves the authenticity of the calculated exponents. 

\par
In table II, we  show the theoretical values of critical exponents for mean field, 3D Ising and 3D Heisenberg models.  Notably, the calculated exponents do not match well with any of the 3D models. Similar discrepancy  was also observed in various other itinerant magnetic systems and it has been attributed to the length scale of interaction. The 3D Ising and Heisenberg models described in table II are of {\it short range} type, {\it i.e.} the spin spin interaction falls off rapidly with distance. However, for itinerant system, the interaction can be of long range due to the mobile electrons. It has been observed that for long range interaction, the spin-spin interaction term varies as $J(r) \sim r^{-(d + \sigma)}$, where $d$ is the effective dimensionality of the system, $r$ is the spin-spin distance and $\sigma$ is an exponent.~\cite{fisher} It has been found that for $\sigma <$ 2,  such model for long range interaction can hold good. 

 \par
 In case Cr-Fe itinerant system~\cite{fecr}, the observed critical exponents were explained on the basis of long range interaction.~\cite{fisher} It was found that for $\sigma$ = 1.34, the observed exponents ($\beta$ =0.298 $\gamma$ =1.392 $\delta$ =5.67 ) match well with  2D Ising model  coupled with long range interaction. Interestingly, like Cr-Fe system, the critical exponents obtained for Y$_2$Ni$_7$ from the present study are also very close to the 2D Ising model with long range interaction. Therefore, it appears  that Y$_2$Ni$_7$ also has the same universality class of Cr-Fe alloys as far as the critical exponents are concerned. Prompted by the similarity, we have also used the same method to calculate the value of $\sigma$ that produces the best suitable set of critical exponents. For long range interaction, the exponent $\gamma = \mathcal{F}\{\sigma, d, n\}$, where $\mathcal{F}$ is a known function (see equation 9 of reference~\cite{fisher}) and $n$ is the dimension of the order parameter. The other exponents are related to $\sigma$ as $\nu = \gamma/\sigma$, $\alpha$ = 2-$\nu d$, $\beta$ = (2$ -\alpha - \gamma$)/2, and $\delta$ = 1 + $\gamma/\beta$. We have used an iterative method to calculate $\sigma$ from these relations starting from the experimental value of $\gamma$ (= 1.40) and the best match to our experimental data is obtained for $d$ = 2, $n$ = 1 and $\sigma$ = 1.38.  This  long range 2D Ising model with $\sigma$ = 1.38 produces critical exponents $\beta$ = 0.314, $\gamma$ = 1.40,  and $\delta$ = 5.46, which match very well with  our experimentally obtained values. It is worth mentioning that recently such long range model for magnetic interaction was also found to be suitable for explaining the critical behavior of itinerant manganite Pr$_{0.5}$Sr$_{0.5}$MnO$_3$.~\cite{pramanik}
\par     
The possible long range spin spin interaction in case of Y$_2$Ni$_7$ can be understood on the itinerant character of the electrons responsible for magnetism. Since the electrons are delocalized, the spin-spin interaction can be of long range. The 2D  Ising Character of the magnetic interaction can have its origin in the anisotropic crystal structure.  Y$_2$Ni$_7$ has  Gd$_2$Co$_7$ type rhombohedral structure, which can be considered to be derived by stacking YNi$_5$ and YNi$_2$ slabs in  2:1 ratio along the $c$ axis. Such layered arrangements of atoms can  give rise to effectively 2D character of the magnetic interaction.

\section{Acknowledgment}
AB wishes to thank Council for Scientific and Industrial Research (CSIR), India for his research fellowship.The authors would like to  acknowledge the Low Temperature \& High Magnetic Field (LTHM) facilities at CSR, Indore (sponsored by DST) for heat capacity measurements.

\end{document}